%% file: main_arXiv.tex
\documentclass[a4paper,11pt]{article}
\usepackage[utf8]{inputenc}
\usepackage[T1]{fontenc}
\usepackage{graphicx,url}
\usepackage{a4wide}
\usepackage{mathpazo}
\usepackage[font=footnotesize,labelfont=bf]{caption}
\usepackage{amsmath}
\usepackage{bm}
\usepackage{amssymb}
\usepackage{multirow}
\usepackage{color,soul}
\usepackage{xurl}
\usepackage{authblk}
\usepackage[hidelinks]{hyperref}
\usepackage{adjustbox}
\usepackage[shortlabels]{enumitem}
\setlist[enumerate]{nosep}
\usepackage[table,xcdraw]{xcolor}
\usepackage{tabulary}
\usepackage{booktabs}
\usepackage{float} 

\usepackage{hyperref}

\definecolor{Magenta}{rgb}{0.8, 0.1, 0.6}

\definecolor{Orange}{rgb}{1, 0.6, 0}

\usepackage{verbatim}

\begin{document}

\title{HoWDe: a validated algorithm for Home and Work location Detection} 
\author[1,*]{Sílvia De Sojo}
\author[2,3,*,$\dagger$]{Lorenzo Lucchini}
\author[3,4]{Ollin D. Langle-Chimal}
\author[3]{\\Samuel P. Fraiberger}
\author[1,$\dagger$]{Laura Alessandretti}

\affil[1]{DTU Compute, Technical University of Denmark, Lyngby, Denmark}
\affil[2]{DONDENA and BIDSA research centres, Bocconi University, Milan, Italy}
\affil[3]{The World Bank 1818 H Street, NW Washington, USA}
\affil[4]{University of California, Berkeley, USA}
\affil[ ]{ }
\affil[*]{\textit{These authors contributed equally to this work.}}
\affil[$\dagger$]{\textit{Corresponding authors: }\texttt{lorenzo.lucchini@unibocconi.it}, \texttt{lauale@dtu.dk}}


\maketitle
\begin{abstract}
\noindent Smartphone location data have become a key resource for understanding urban mobility, yet extracting actionable insights requires robust and reproducible preprocessing pipelines. 
A central step is the identification of individuals’ home and work locations, which underpins analyses of commuting, employment, accessibility, and socioeconomic patterns. 
However, existing approaches are often ad hoc, data-specific, and difficult to reproduce, limiting comparability across studies and datasets.
We introduce \textit{HoWDe}, an open-source software library for detecting home and work locations from large-scale mobility data. \textit{HoWDe} implements a transparent, modular pipeline explicitly designed to handle missing data, heterogeneous sampling rates, and differences in data sparsity across individuals. 
The code allows users to tune a small set of interpretable parameters, enabling to adapt the algorithm to diverse applications and datasets.
Using two unique ground truth datasets comprising $5{,}099$ individuals across $68$ countries, we show that \textit{HoWDe} achieves home and work detection accuracies of up to $97\%$ and $88\%$, respectively, with consistent performance across demographic groups and geographic contexts. We further demonstrate how parameter settings propagate to downstream metrics such as employment estimates and commuting flows, highlighting the importance of transparent methodological choices.
By providing a validated, documented, and easily deployable pipeline, \textit{HoWDe} supports scalable in-house preprocessing and facilitates the sharing of privacy-preserving mobility datasets. Our software and evaluation benchmarks establish methodological standards that enhance the robustness and reproducibility of human mobility research at urban and national scales. \\


\end{abstract}


\section{Introduction} 
Location data collected from smartphones has become a key resource for the analysis of urban systems and human mobility~\cite{gonzalez2008understanding, oliver2020mobile, moro2021mobility, rout2021using, dong2024defining, liu2021identifying}. 
These datasets enable the study of large, diverse populations and have been used within a range of areas, including the study of human movements~\cite{alessandretti2020scales,Centellegher2025,alessandretti2018evidence}, epidemic spreading~\cite{viboud2006predicting,bosetti2020heterogeneity,aleta2022quantifying,alessandretti2022human}, migration patterns~\cite{rango2016data,bosetti2020heterogeneity}, transportation modeling~\cite{mazloumi2010using}, urban planning~\cite{zheng2014urban, rout2021using, dong2024defining, liu2021identifying}, and assessing the impact of natural disasters and societal shocks on population movements~\cite{wilson2016rapid,yabe2022toward,lucchini2021living,lucchini2023socioeconomic,Centellegher2025}.
As the availability of such data increases,  developing robust methods for location data analysis has become a pressing need for scientists and practitioners~\cite{barreras2024exciting}. 

A critical component of location data analyses is the detection of individuals' residential~\cite{2020USCensus,2024Eurostat,2024India}, and work locations~\cite{barreras2024exciting}. 
The accurate identification of home and work locations is key for mobility research, as it enables the analysis of factors such as socio-economic background~\cite{frias2010towards,rowangould2013using,lucchini2023socioeconomic}, commuting behaviors~\cite{calabrese2011estimating,Centellegher2025}, social contact networks~\cite{eagle2009inferring,lucchini2021living},
and accessibility to amenities and services~\cite{gao2015spatio,lucchini2021living}.
More broadly, it contributes to advancements in computational demography and social science research~\cite{pappalardo2023future,barreras2024exciting}.

The most widely used methods for home and work detection are heuristic approaches that leverage the regularity of human behaviour, such as the tendency to spend nights at home and working hours at a workplace~\cite{ashbrook2003using,TheAtlasofInequality,jiang2016timegeo}.
However, these methods face significant challenges.
First, they are sensitive to missing data, which often stems from inactivity or poor signal and can bias home and work detection~\cite{wesolowski2012heterogeneous, wesolowski2013impact, barreras2024exciting}.
Second, they rely on the choice of parameters, such as the time span required to establish a home location or the minimum number of visits required to identify a valid work location.
These limitations underscore the need for methods that systematically address both the heterogeneity of missing data within individuals and the sensitivity to heuristic parameter definitions~\cite{barbosa2018human, barreras2024exciting}.

Despite their importance for mobility analyses, the extent to which heuristic approaches for home/work location detection are reliable remains insufficiently validated, and there is no consensus on how to choose one specific algorithm and its parameters' values~\cite{frias2010towards, ahas2010using, vanhoof2018assessing, vanhoof2020performance, pappalardo2021evaluation, barreras2024exciting}, in part due to a scarcity of ground truth data. 
Existing validation efforts have been limited to small samples, short time periods, low spatial resolution, or to home detection alone~\cite{vanhoof2020performance, rinzivillo2014purpose, pappalardo2021evaluation, ahas2010using, frias2010towards}.
Furthermore, the influence of key parameter choices (such as the length of the period considered or the minimum activity level necessary to define home/work locations) has not been studied systematically.
Limited validation has hindered the refinement and harmonisation of these methods into standardised best practices ~\cite{barreras2024exciting}. 
In turn, the absence of standard algorithms hinders the comparability and reproducibility of mobility studies ~\cite{jiang2012clustering, barreras2024exciting}.

In this study, we leverage two distinct ground truth datasets collected across more than 68 countries to address these key challenges in home/work location detection. 
Both datasets include sequences of visits to detailed locations with corresponding ground truth: the first dataset (D1) contains self-reported home and work locations from $4,599$ users collected between 2015 and 2019, while the second (D2) is based on independent expert annotations for $500$ users throughout 2020 and 2021.

First, we formalise \textit{HoWDe}, an algorithm for home and work detection that enhances existing heuristics by integrating key elements from previous work into a comprehensive approach.
\textit{HoWDe} combines the explicit detection of both home and work locations~\cite{lucchini2021living,jiang2016timegeo}, the identification of the workplace based on recurrent visits during working hours~\cite{jiang2016timegeo}, the identification of the home as the most frequently visited location at night~\cite{pappalardo2021evaluation, TheAtlasofInequality, vanhoof2020performance, frias2010towards}, a sliding window to identify home and job changes over time~\cite{lucchini2021living}, with tunable window length~\cite{lucchini2021living}. 
Additionally, \textit{HoWDe} improves upon existing approaches by accommodating users whose data sparsity varies over time.
This is achieved by identifying home and work locations based on the proportion of time spent at a given location, while controlling for individual-level sparsity rather than excluding users based on absolute time or visit thresholds, as in prior work.

Second, we assess \textit{HoWDe}'s performance against the ground truth data. 
We provide a detailed analysis of the role of its parameters, and we study how performance changes across age groups, gender, countries, and urbanisation levels.
We quantify how errors in the estimation of home and work locations propagate to key downstream tasks such as estimating employment rates or commuting distances.

The article is organized as follows: in Section \ref{sec:lit_rev} we introduce existing approaches to home and work location detection; 
in Section \ref{sec:mat_met} we introduce the ground truth data and additional methods used in this study; 
in Section \ref{sec:temppatt}
we explore the ground truth data to motivate the heuristic choices of \textit{HoWDe},
in Section \ref{sec:algorithms} we describe the \textit{HoWDe} algorithm; in Section \ref{sec:results} we evaluate the performance of the algorithm, explore the effect of parameter choices, assess robustness across demographic groups, and demonstrate applications in estimating employment rates and commuting patterns.


\section{Background}\label{sec:lit_rev}
Smartphones use an in-built GPS sensor to estimate the device's latitudes and longitude coordinates over time, at a frequency that varies depending on device usage.  
The device position is then collected by various location-based apps, such as navigation or social media apps.
These data can be used by location intelligence companies or researchers to study human mobility and its regularities, both at the individual and at the population level.
Raw location data utilised in urban and mobility studies consists of timestamped sequences of \textit{(latitude, longitude)} records associated with a unique user ID. 

The first pre-processing step involved in the analysis is the so-called ``stop-location detection"~\cite{aslak2020infostop,hariharan2004project}. 
Stop location detection consists of (i) identifying records that are captured when individuals stay within a small area for a sufficient amount of time (e.g. when they are at home, at the office, or in a restaurant); (ii) applying spatial clustering to all these ``static records", to identify a set of unique locations, i.e. those places where people spend time.
Stop location definition might vary depending on the type of data analyzed and the specific characteristics of the data points, e.g., whether you use Call Detail Records (CDR), eXtended Detail Records (XDR), or GPS points. 
However, the principled choices these algorithms are founded on can naturally be extended to any of these different cases, making them easily generalizable. In this work, stop locations are defined as spatiotemporally close sets of GPS points. For further details, we refer the reader to the Supplementary material (Section S-2) and prior work~\cite{aslak2020infostop,lucchini2021living}.

After detecting the individuals' stop locations, individual trajectories can be summarised as sequences of stays.
Each stay is associated with a start time (when the stay started), a duration (how long the stay lasted), and a location ID (where the stay happened). 

A common next step in many mobility studies involves identifying key locations, such as individuals’ homes and workplaces.
To achieve this, researchers and practitioners typically assume that visits to home and work locations exhibit specific daily and weekly regularities. 
There are various approaches to identifying home and work locations. 
The majority of studies adopt ad-hoc heuristics~\cite{ahas2010using,frias2010towards,vanhoof2018assessing,pappalardo2021evaluation,jiang2016timegeo, tozluoğlu2024mobilephoneapplicationdata,verma2024comparison}. 
The most commonly used heuristics rely on rules in relation to ``how much'', and ``when'' an individual visits each location during the period of observation~\cite{ahas2010using,frias2010towards,vanhoof2018assessing, csaji2013exploring, phithakkitnukoon2012socio,chen2021identifying}. 
Some studies complements temporal heuristics with semantic information, for example by matching stay locations to nearby points of interest (POIs) or land-use categories to ensure that inferred homes fall within residential areas and work locations within commercial or industrial zones~\cite{isaacman2011identifying,hasan2013understanding}.
Importantly, in most cases, these heuristics are not validated against ground truth data nor are standardised across studies. 
A minority of studies, instead, leverage machine learning techniques~\cite{liao2007extracting, rinzivillo2014purpose}; however, these methods either generalise poorly to individuals not seen during training~\cite{liao2007extracting}, or have been tested only in specific cities and rely on a particular mode of transport, such as private cars~\cite{rinzivillo2014purpose}.

In this work, we compare our algorithm against two established baseline methods commonly used in mobility research, selected for their widespread adoption and their relevance to circadian and regularity-based home-work detection heuristics.
These methods were designed and utilised over the past decade, leveraging diverse sets of mobile-phone-based data (e.g., Call Detail Records (CDRs), eXtended Detail Records, (XDRs), Control Plane Records (CPRs), or GPS records)~\cite{frias2010towards,jo2012spatiotemporal,jiang2017activity,pappalardo2021evaluation} as well as activity-based travel surveys~\cite{jiang2012clustering}.
Among the home/work detection approaches in the literature, these are the most widely used with GPS data because they rely solely on circadian regularities, making them easy to deploy across heterogeneous mobility datasets without requiring POIs, land-use information, or detailed ground truth. Both approaches build on the rationale of ranking locations based on the day of the week (weekdays vs. weekends) and the time of day (daytime vs. nighttime) when visits occur.
In the following, we describe these two methods in detail. 

\subsection{The \textit{Atlas} approach}
The method used in~\cite{moro2021mobility, TheAtlasofInequality, moro2023diversity} (hereinafter \textit{Atlas} method) defines a home location as the most probable location visited between 10 p.m. and 6 a.m., i.e., the stop location where a device was found spending most of the nighttime.
The scikit-mobility Python library—a standard toolkit for mobility analysis—implements this definition~\cite{pappalardo2022scikit}.

To ensure that only users with sufficient data quality are included in the analysis, a filtering strategy is typically applied. 
Individuals who spend fewer than 10 nights in the same census block group during a six-month observation period~\cite{moro2021mobility} (or fewer than 5 nights over a three-month period~\cite{moro2023diversity}) are excluded from the analysis.
The \textit{Atlas} method does not explicitly estimate work locations; however, its approach to identifying home locations can be extended to infer work. 
Following a similar rationale, we define an \textit{Atlas} work location as the most common location on weekdays between 9 a.m. and 4 p.m.

\subsection{The \textit{TimeGeo} approach}
More complex heuristics are adopted in~\cite{jiang2016timegeo} (hereinafter referred to as the \textit{TimeGeo} method), in particular for identifying work locations. 
\textit{TimeGeo} ranks stop locations based on the frequency of visits during weekday nights (between 7 p.m. and 8 a.m.) and weekends and picks the most visited location as an individual's home. 
The same logic is applied when detecting the work location, but now focusing on weekday work time periods (between 8 a.m. and 7 p.m.) and including additional requirements:
i) the candidate work location is visited at least three times during the study period, and ii) the distance between the candidate work location and the individual’s home location exceeds 500 meters. 
Similar to the \textit{Atlas} method, \textit{TimGeo} applies a data quality filter, excluding users with fewer than 50 total stops and fewer than 10 home stays during the observation period.

\paragraph*{}
\textit{Atlas} and \textit{TimeGeo} rely on similar strategies for handling missing data. 
After estimating home and work locations, users are filtered based on their overall activity levels and/or the number of nights spent at home. 
While this filtering excludes individuals with low mobility or no stable residence, it has key limitations.
First, existing methods fail to account for temporal heterogeneity in data availability within individuals. 
Gaps in location data—caused by device inactivity, disabled services, or poor signal—are common in smartphone records and vary widely across users and within users over time ~\cite{wesolowski2012quantifying, wesolowski2013impact, barreras2024exciting}. These inconsistencies can introduce systematic biases in the inference of home and work locations, particularly when left unaccounted for~\cite{barreras2024exciting}.
Second, it relies on parameter choices that are typically ad hoc, lacking empirical grounding, and are often based on visit counts, a design that can disproportionately impact users with shorter or less consistent observation periods.

Moreover, systematic large-scale validation using GPS data remains limited~\cite{barreras2024exciting}. 
Most existing evaluations rely on lower-resolution sources such as CDRs, XDRs, and CPRs~\cite{pappalardo2021evaluation}, which constrain temporal precision and spatial accuracy.

\textit{HoWDe} addresses these gaps by grounding its design in large-scale GPS data annotated with home and work locations. The methodology we propose integrates established processing techniques while explicitly addressing the challenges of data sparsity and parameter sensitivity, offering a standardised and scalable approach to home–work location detection.

\section{Materials and Methods}\label{sec:mat_met}
\subsection{Ground truth datasets\label{sec:data}}
To validate the proposed methods, we use two GPS-based datasets with ground truth home and work labels: one featuring large-scale self-reported annotations, and the other comprising expert-labelled trajectories. 
Together, they enable robust evaluation across diverse populations and settings.

\subsubsection{Dataset 1: DTU stop data and self-reported annotations}\label{subsec:D1}
This dataset uses mobility data collected between 2017 and 2019, previously analyzed in~\cite{alessandretti2020scales}, where stop locations were identified using the scalable stop-location detection algorithm InfoStop~\cite{aslak2020infostop}. 
The stop location data captures the mobility patterns of 96,065 users aged 25 to 64 years old.

The dataset includes $2,309$ self-reported home and $2,479$ work location labels, collected from $4,599$ mobile phone and smart band users from 68 countries across continents. 
The labelled sample is biased towards urban areas and higher- to middle-income countries. 
To preserve privacy, spatial coordinates have been removed (See Section~\ref{sec:data_code_avail} for data availability).

\subsubsection{Dataset 2: Veraset Movements data and annotation}\label{subsec:D2}
This dataset is based on anonymised GPS location data provided by \textit{Veraset} as part of the \textit{World Bank’s} “Monitoring COVID-19 Policy Response through Human Mobility Data” project.
Unlike most GPS providers, \textit{Veraset} sources its data from thousands of Software Development Kits (SDKs), reducing sampling bias and covering approximately $5\%$ of the global population~\cite{VerasetMovementVeraset}.
The data spans 24 months, from January 1, 2020, to December 31, 2021, and has been previously used to assess the impact of the pandemic and the policy responses~\cite{lucchini2023socioeconomic}.

Dataset D2 includes GPS trajectories from $100$ devices in each of five countries: Brazil, Colombia, Indonesia, Mexico, and the Philippines. For each country, users were selected to reflect high-, medium-, and low-wealth neighbourhoods ($33\%$), based on a wealth classification framework developed for GPS mobility analysis in middle-income countries~\cite{lucchini2023socioeconomic}.
To ensure data quality, users were required to have stop records on at least 20\% of days both before the WHO’s pandemic declaration and over the 2020 pandemic period.

For each user, experts annotated the stop location sequences for 2020 by inspecting location visits, satellite imagery, and local amenities (see Supplementary material Section S-1 for a detailed description of the annotation procedure). 
Throughout the analysis, we extended these annotations to 2021 data to increase the amount of available location data for the selected users. We note that data for 2021 contains substantially fewer observations and exhibits marked shifts in data quality associated with pandemic-related changes in mobility behaviour.

The final sample includes 500 users, of whom $287$ have at least one work location labelled.
Since annotators inspected the complete longitudinal stop sequence, it is not possible to explicitly consider changes in home and work for each individual.

\subsection{Measuring ground truth home/work visits to inform \textit{HoWDe} design\label{sec:additional_methods}}
In Section \ref{sec:temppatt}, we examine empirical regularities in home and work visits using the ground truth data.
Here, we detail the methods used to analyse these patterns. 
First, we outline how we identified typical daily patterns of home and work visits, which we refer to as \textit{behavioural clusters}.
Second, we describe how we assessed whether individuals in our dataset are captured by a single behavioural cluster or distributed across multiple clusters. 

\subsubsection{Home and work visits behavioural clusters}\label{subsec:HW_visits_clustering}

To characterise daily regularities in home and work visits, we represent each user’s behaviour as a sequence of 24 hourly bins, corresponding to one day.
Each hourly bin is assigned one of three categories: (1) visit to the home (or work) location, (2) a visit to another location (e.g., work, leisure, or other activities), and (3) missing data.
For home patterns, category~(1) corresponds to being at the home location; for work patterns, it corresponds to being at the workplace, while hours spent at home are labelled as visits to another location (2).

These daily sequences capture the temporal structure of individuals’ visits, allowing us to quantify variation in visit regularity across users and days.
We cluster these sequences using the K-Modes algorithm—a categorical adaptation of K-Means~\cite{cao2009new}—to identify days exhibiting similar visitation patterns, which we refer to as behavioural clusters.
The number of clusters is determined using the elbow method.
Rather than defining a strict typology, here our aim is to uncover dominant behavioural profiles that summarise the most common home and work visitation patterns (see Section~\ref{sec:temppatt}).
These clusters form the basis for assessing whether individuals’ behaviour aligns with a single dominant pattern or spans multiple profiles, informing the design of \textit{HoWDe}.

\subsubsection{Behavioral profile entropy}\label{sec:behprof_entropy}
To quantify how consistently users follow specific behavioural profiles, we measure the normalised Shannon entropy \cite{cover1991entropy}, treating each behavioural cluster as a distinct state.
The entropy is defined as:
\begin{equation}
H_{\text{norm}} = - \frac{1}{\log{k}}\sum_{i=1}^{k} p_i \log{p_i},
\end{equation}
where $k$ is the total number of possible clusters, and $p_i = \frac{n_i}{N}$ denotes the probability of a user being assigned to cluster $i$, with $n_i$ being the number of days classified in that cluster, and $N$ the total number of days for the user. 

Normalising by the maximum entropy $\frac{1}{\log{k}}$ ensures comparability across different settings—such as weekdays vs. weekends, or home vs. work patterns, by bounding the entropy between 0 (only one profile used) to 1 (maximum diversity; multiple profiles and all used equally).

\subsection{Performance evaluation metrics}\label{sec:methods-performance}
In this work, we report the performance of the algorithm in terms of: ``detected accuracy'' and ``fraction of not detected''.
Below, we define these metrics in the context of home detection; the same definitions apply analogously to work detection.

The fraction of `not detected' indicates cases where a home location was labelled, but the algorithm couldn’t assign a home due to thresholds on data quality or a minimum fraction of visits. 

In the case of D1, the dataset consists of a set of user-weeks, corresponding to users who self-reported living in a specific home (the ``ground truth home") in a given week. 
For some of these user-weeks, our algorithm identifies a ``detected home".
We evaluate the performance of the algorithm by comparing pairs of ``ground truth home" and ``detected home" across user-weeks.
The ``fraction of not detected'' is thus computed as $f_{ND} = 1-n_{\bar{H}}/n_{H}$, where $n_{\bar{H}}$ is the number of user-week pairs with a detected home, and $n_{H}$ is the number of user-week pairs with a ground truth home. 
In the case of D2,  where ground truth information does not refer to a specific week, $n_{\bar{H}}$ corresponds to the number of users with a detected home, and $n_{H}$ corresponds to the number of users with a ground truth home. 
In state-of-the-art methods, such as \textit{Atlas} and \textit{TimeGeo}, home locations are always assigned to users who are included in the analysis, resulting in no undetected cases.
However, this practice can introduce systematic errors when data coverage biases are present throughout the study period, e.g., due to data-provider issues, external shock events, or individual changes in activity levels.

The ``detected accuracy'' captures the fraction of correctly labelled home locations among the set of detected home locations, following the approach previously applied by \cite{pappalardo2021evaluation}. 
Therefore, we define detected accuracy as $Acc= m_{\bar{H}}/n_{\bar{H}}$, where $m_{\bar{H}}$ is the number of user-weeks (for D1) or users (for D2) for which the detected home location matches the ground truth home location. 
In dataset D2, if annotators identified multiple homes for an individual, we considered it a match if the individual's detected home location corresponded to at least one of the annotated ground truth home locations. 
Note that our definition does not correspond exactly to the traditional definition of accuracy because we separately assess the algorithm performance in terms of the detection/non-detection of a home (captured by $f_{ND}$) and the detection of the correct home (captured by $Acc$).

The definitions above are also used in the case of work detection, where only users with a reported work location, $n_{W}$, are included in the denominator of $f_{ND}$, and only those with a detected and a matching work location are counted in the nominator and in the detected accuracy computation, as $n_{\bar{W}}$ and $m_{\bar{W}}$, respectively.

\subsection{Comparing home/work detection methodologies}\label{sec:methods-comparison}
We compare the performance of three home/work detection methods, \textit{HoWDe}, \textit{Atlas}, and \textit{TimeGeo} (see Section~\ref{sec:lit_rev}), within a shared evaluation framework.
All methods are applied to an identical set of stop locations, generated using the same processing pipeline for each dataset (see Supplementary material, Section S-2).

As described in Section~\ref{sec:lit_rev}, both \textit{Atlas} and \textit{TimeGeo} typically apply pre-filtering steps before location detection.
Although \textit{HoWDe} includes an internal data-quality assessment, we pre-filter users for all methods to ensure comparability.
Specifically, we retain only users with at least 10 days of data in dataset D1 (note that $83\%$ of these users have $\geq 60$ days) and at least 71 days in dataset D2 (applicable to all D2 users).
This filtering step ensures that performance differences reflect methodological rather than data-quality variations.

To further isolate methodological differences from temporal definitions, we adopt consistent nighttime (home) and daytime (work) windows across all methods, preserving the weekend–weekday distinctions defined in the baselines (see Section~\ref{sec:lit_rev}).
These time ranges are empirically determined using ground truth data (see Section~\ref{sec:temppatt}).


\section{Empirical regularities in home and work visits}\label{sec:temppatt}
To establish validated standards for home and work location detection, we first examine the temporal patterns that characterise visits to home and work locations in our ground truth dataset.
The \textit{HoWDe} algorithm (presented in Section \ref{sec:algorithms} below) is built on these insights.

\begin{figure}[!ht]
\centering
\includegraphics[width=1\textwidth]{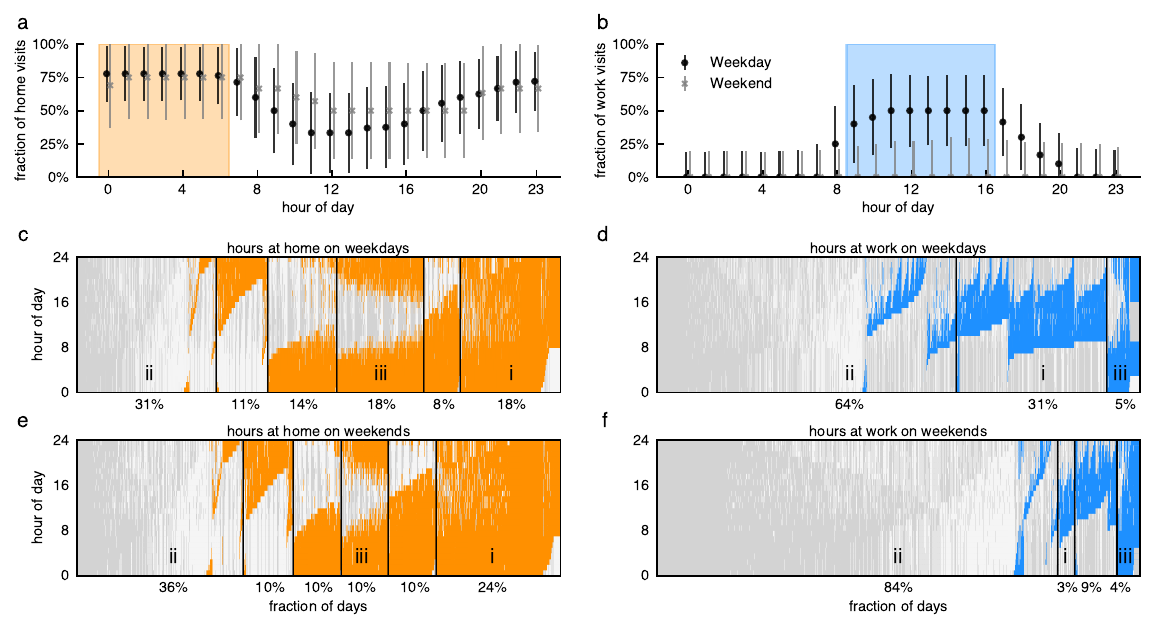}
\caption{\small \textbf{Behavioral visit profiles to home and work locations}
    a. Fraction of visits per hour of the day to the annotated home locations in dataset D1, for weekdays (black) and weekends (grey) across individuals. The shaded orange area highlights the period with the highest probability of being at home.
    b. Fraction of visits per hour of the day to the annotated work locations in dataset 1, for weekdays and weekends. The shaded blue area highlights the period with the highest probability of being at work during weekdays.
    c. User-day profiles by hours at home (orange), other locations (grey), and periods without location data (light-grey) during weekdays. 
    d. Hours at work (blue), other location (grey), or without data (light-grey), per hour of the day across days with data for weekdays.
    e-f. Show user-day profiles during weekends for hours at home (orange) and work (blue) respectively. 
    The user-day profiles are sorted by cluster, hour of the first home visit, total time spent at home, and hours without data. Clusters are separated by black vertical lines, labelled with roman numerals, and the fraction of days per cluster is noted on the x-axis.
    }
\label{fig:home_work_patterns}
\end{figure}

We analyse the regularity of home and work visits by measuring how often and when individuals are recorded at their annotated location (see Section~\ref{sec:data}).
Figure~\ref{fig:home_work_patterns}a-b shows the probability of being at home or work by hour of the day.
We find that home visits peak between midnight and 6 a.m., while work visits are most frequent during typical business hours (between 9 a.m. and 4 p.m.).

Patterns of visits to the home locations are similar between weekdays and weekends (see Figure~\ref{fig:home_work_patterns}a). 
For the work location, we observe a drastic reduction in the fraction of individual visiting their work location during weekends (see Figure~\ref{fig:home_work_patterns}b).

Results are consistent across both datasets, with overall higher probabilities of being at home (and lower probabilities of being at work) found in D2 due to the COVID-19 pandemic containment policies in place during most of the period covered by the data (see Supplementary material Sec. S-3).
 
Our analysis suggests that work location is best determined using weekday daytime hours, while home location can be identified using night hours, regardless of weekdays or weekends, similar to previous approaches~\cite{moro2021mobility,TheAtlasofInequality, pappalardo2021evaluation}.

However, while general trends exist, we also find significant variations across days and users, highlighting pronounced behavioural heterogeneities at the individual level (Figure~\ref{fig:home_work_patterns}a-b).
Figure~\ref{fig:home_work_patterns} panels c-f report disaggregated visit patterns for ``user-day'' pairs, where each vertical line represents one person’s hourly sequence of visits to the home or work location over a single day. 
Most individuals’ daily patterns can be grouped into distinct behavioural profiles.
We identify six profiles based on home visits and three based on workday visits (see Section~\ref{subsec:HW_visits_clustering} for details on how behavioural profiles are obtained).

With respect to home visits, three profiles account for most user-days:
i) days when users mostly stay at home, ii) when they are rarely/never at home, and iii) when they are away from home during typical business hours (see Figure~\ref{fig:home_work_patterns}c,e).
Similarly, work-behaviour profiles are described by three main daily patterns: i) days when users spend business hours at work, ii) when they are rarely/never at work, and iii) when they spend night hours at work (Figure~\ref{fig:home_work_patterns}d,f). 
We find that the distribution of individuals across these profiles is generally independent of their socio-demographic characteristics, such as age, gender, and urbanisation (see Supplementary material Sec.S-4). 

As the behavioural profiles we identified are based on single user-day activity data, an important question is whether individuals switch between these profiles or, rather, a single profile consistently represents them. 
We find that users visit a median of $5$ out of $6$ home clusters and $2$ out of $3$ work clusters (for both D1 and D2). 
We measure the individual inter-profile variability through the normalised entropy, a measure ranging from $0$ (when all days of a user fall into a single profile) to $1$ (when all profiles are equally important) \cite{cover1991entropy} (see Sec.~\ref{sec:behprof_entropy}).
On average, users exhibit a normalised entropy of $0.66 \pm 0.01$ for home-profile changes between weekdays and $0.48 \pm 0.01$ for weekends ($0.62 \pm 0.01$ and $0.42 \pm 0.01$ for D2, respectively), and $0.51\pm 0.01$ for work-profile changes on weekdays only ($0.40 \pm 0.02$ for D2). 
These values are comparable to the ones obtained when randomly allocating profiles among user days, confirming that users do not have a single dominant behavioural profile (see Table S-1, in Supplementary material Sec. S-5)..

All in all, our findings suggest that focusing on the same profiles across users is unlikely to favour one specific socio-demographic group, thus supporting the choice of focusing on weekday daytime hours (9 a.m. to 4 p.m.) for detecting work locations and nighttime hours (12 a.m. to 6 a.m.) for detecting home locations, regardless of weekdays and weekends. 
These time windows, combined with minimum data coverage requirements, form the foundation of the heuristics implemented by the \textit{HoWDe} algorithm.

\section{\textit{HoWDe}: a Home and Work Detection algorithm}\label{sec:algorithms}
In this work, we introduce \textit{HoWDe}, a dynamic home and work detection algorithm that captures residential moves and job changes over time using a sliding window approach. 
The heuristics adopted by \textit{HoWDe} build on the empirical results obtained in the previous Section \ref{sec:temppatt}, leveraging our comprehensive ground truth data.

A key feature of \textit{HoWDe} is its ability to handle data sparsity through a series of internal steps designed to ensure robust home and work detection even when individual traces are incomplete.
Rather than relying on absolute time spent at a location~\cite{jiang2016timegeo,TheAtlasofInequality,lucchini2021living,lucchini2023socioeconomic,vanhoof2018assessing,vanhoof2020performance}, \textit{HoWDe} identifies locations based on the proportion of time spent, a choice that naturally accommodates users with uneven data sparsity over time (see Section~\ref{sec:temppatt}). 
This proportional representation is combined with fixed-length hourly sequences and tunable temporal coverage filters, allowing the algorithm to standardize temporal structure across days while explicitly accounting for missing data. 
Together, these design choices enable HoWDe to remain robust under varying levels of sparsity both across and within individuals.

We provide an open-source implementation of \textit{HoWDe} in PySpark \cite{gitRepoCode}.
Supplementary materials section. S-6 outlines all aspects of the software implementation, including availability, packaging, licensing, engineering decisions, and the development model. 
We further emphasise computational efficiency in Sec. S-6.4, where we compare the resource utilisation of \textit{HoWDe} with that of the benchmark methodologies (\textit{Atlas} and \textit{TimeGeo}).

The next subsections detail the algorithm workflow (Section~\ref{sec:howde-overview}) and its tunable parameters (Section~\ref{sec:howde_params}). 

\subsection{\textit{HoWDe} workflow overview}\label{sec:howde-overview}
\textit{HoWDe} takes as input a user's pre-processed sequence of stops, where each stop is fully defined by a user-location unique ID and a start and end timestamp. 
First, we perform a temporal aggregation of the stop sequences into hourly-bin sequences. 
In each hourly bin, we retain the location where the individual spent the most time.
This aggregation limits sequences to a maximum of 24 locations per day, boosting computational efficiency and making \textit{HoWDe} faster than previous approaches based on sliding windows~\cite{lucchini2021living,lucchini2023socioeconomic}, with only limited information loss. 
An extended evaluation of how increasing temporal resolution beyond 1-hour bins affects performance is provided in Supplementary materials sec. S-7.

\begin{figure}[h!]
    \centering
    \includegraphics[width=1\textwidth]{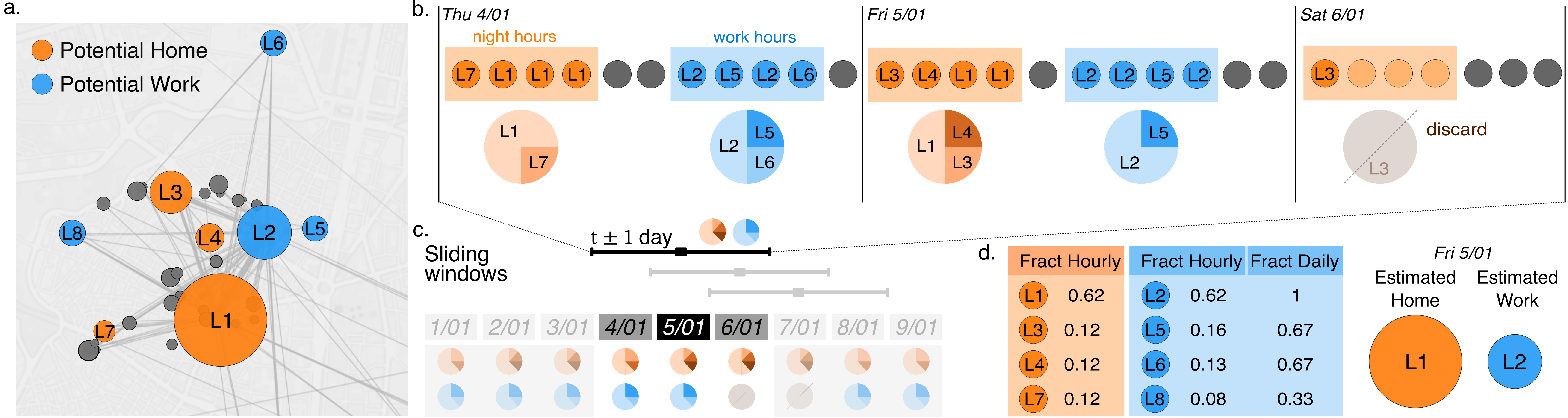}
    \caption{\small \textbf{Schematics of the \textit{HoWDe} algorithm workflow for home/work location detection.} 
    a. Network of visited locations. The diameter represents the number of visits. In orange, we report the stops visited during night hours, and in blue are the ones visited during work hours.
    b. Sequences of locations visited hourly. The pie charts display the allocation of visits across locations during the night hours (orange) and work hours (blue) on each day.
    Note that on Saturday, 6/01, the night hours data was discarded since there were not enough hours with available data (exemplifying the temporal coverage filter tuned by the parameter $C_{hours}$). 
    In this minimal example, no location appears during both night and work hours; however, this may happen in real data.
    c. For each day, $t$ we illustrate a sliding window of $3$ consecutive days centered on $t$. 
    For work detection, the sliding window excludes weekends. 
    d. Aggregation step for the window centred on Friday, 5th. For each location, we compute the average fraction of time it was visited during night hours (orange) and during work hours (blue). Additionally, we compute the fraction of days the location is visited at least once during work hours. Locations are then sorted in descending order by on these fractions, and the top ones are selected as the estimated Home (L1) and Work (L2) locations for Friday.}
    \label{fig:algorithm_desc}
\end{figure}

Second, we exclude days with insufficient hourly bins with data (this step is controlled by the tunable parameter $C_{hours}$, see Table \ref{tab:parameters}).
This step accounts for the observation that the amount of missing data varies widely across users and over time, and even within individual users’ routines (see Section~\ref{sec:temppatt}).
This is done separately for work detection (where we consider business hours) and home detection (where we consider night hours, see Figure~\ref{fig:algorithm_desc}b).
Hence, it is possible that one day is retained only for home detection or for work detection. 

Third, for each day $t$, we calculate the \textit{fraction of hourly-bins} spent at each location during nighttime (midnight to 6 a.m. for home detection), and typical business hours (9 a.m. to 4 p.m. on weekdays, for work detection). 
This choice is motivated by the observation that individuals are most likely at home between midnight and 6 a.m. ($\sim75\%$ probability) and at work between 9 a.m. and 4 p.m. on workdays ($\sim50\%$), see Section \ref{sec:temppatt}. 
To compute such a fraction, we use as a denominator the number of hourly bins with records (Figure~\ref{fig:algorithm_desc}.b).
Unlike previous approaches that rely on absolute time spent at each location, considering fractions can better accommodate users with varying amounts of missing data over time.

Fourth, for each day $t$, we consider a \textit{window} of size $\Delta_T + 1$, covering the range included in $t\pm \Delta_T/2$ days (see Figure\ref{fig:algorithm_desc}.c). 
A configurable parameter allows restricting the window to past data only, corresponding to the interval $t -\Delta_T$ days (see Supplementary materials Sec. S-8).
This window acts as a memory component for the algorithm to better detect behaviour in the period before and after a specific day.
In each window, we average the fraction of hours a location was visited (see Figure \ref{fig:algorithm_desc}c).

Thus, for each window, we obtain a collection of potential home and work locations. 
Among these, we retain only those that meet minimum requirements with respect to the proportion of visits within and across days. 
These requirements are controlled by parameters $C_{days}$, $f_{hours}$, $f_{days}$ (see Table \ref{tab:parameters}).
If no location meets the requirements, no home (or work) is assigned for that specific window.

Finally, we sort the home and work collections in two different ways. 
For home, we sort based on the typical fraction of hourly bins the location is visited within a day. 
For work, we sort based on the typical fraction of days the location is visited within the window. 
If no work location can be selected based on the fraction of days, the sorting is based on the fraction of hourly bins. 
The top location from these two lists is then assigned as home (or work). 
By prioritising the fraction of days over the fraction of hours for work detection, we are emphasising work as a location defined by regular visits throughout the workweek (see Supplementary materials Sec. S-8).

In summary, the algorithm offers control over two key aspects. 
It allows users to define the threshold for determining when there is enough data to attempt home or work detection.
It also enables users to set the criteria for determining whether a location qualifies as a candidate home or work site, while accounting for individual differences in data completeness.

\subsection{\textit{HoWDe} configurable parameters}\label{sec:howde_params}
To adapt to diverse datasets and research goals, \textit{HoWDe} includes a set of tunable parameters that govern the algorithm's temporal coverage filtering and the temporal consistency required to assign home and work locations. 

First, we define the size of the memory window to be configured separately for home and work $\Delta \mathbf{T_H}$, $\Delta \mathbf{T_W}$.

Second, the coverage parameters specify the minimum temporal coverage required at the daily ($C_{hours}$) and the window level ($C_{days,H}$, $C_{days,W}$):

\begin{itemize}
    \item[] $\mathbf{C_{hours}}$:  This parameter specifies the minimum fraction of night- or business-hour bins that must contain data for a day to be considered in the detection process. By default, \textit{HoWDe} retains only days with at least 40\% of the hourly bins containing stop information, ensuring a minimum level of temporal coverage. The threshold can be adjusted depending on the data quality or the application context.

    \item[] \textbf{$\mathbf{C_{days,H}}$ and $\mathbf{C_{days,W}}$}: To account for variability in data completeness across datasets, \textit{HoWDe} requires a minimum \emph{fraction of days with data} within each sliding window. These parameters independently control home and work detection. If the fraction of days with data in the window falls below the threshold, the algorithm marks the day as having an “undetected” home or work location.

\end{itemize}

Third, the location-selection parameters determine when a place qualifies as a potential home or work location, based on the minimum fraction of hours within a day ($f_{hours,H}$, $f_{hours,W}$) and the minimum fraction of days within a window ($f_{days,W}$):

\begin{itemize}
    \item[] \textbf{$\mathbf{f_{hours,H}}$ and $\mathbf{f_{hours,W}}$}: These parameters set the minimum average \emph{fraction of hourly bins} (night or business hours, respectively) during which a candidate location must be visited to qualify as home or work. If the location’s average fraction of hourly-bin visits over the window falls below this threshold, it is excluded as a potential home or work location. 

    \item[]  \textbf{$\mathbf{f_{days,W}}$}: Work locations often follow different visit regularities than home locations, reflecting local labour norms (e.g., four-day work weeks). To capture these patterns, \textit{HoWDe} requires that a work location be visited on at least a minimum fraction of days with data within the window. This parameter does not apply to home detection, which typically assumes near-daily presence when data are available.
\end{itemize}

\begin{table}[H]
    \centering
    \renewcommand{\arraystretch}{1.2} 
    \setlength{\tabcolsep}{8pt}     
    \caption{\textbf{Summary description of the \textit{HoWDe} configurable parameters.} 
    The table reports the parameter names, whether they are used in home detection, work detection, or both, and the description of the parameter's purpose. 
    When looking at the hourly bins with data (hours with data), that is the case of $C_{hours}$, $f_{hours,H}$, and $f_{hours,W}$, fractions are computed with respect to night/business hourly bins only. Note that $C_{hours}$, $C_{days,H}$, and $C_{days,W}$ act as temporal coverage filters to ensure a given day has enough data to estimate home or work location. 
    Instead, $f_{hours,H}$, $f_{hours,W}$, and $f_{days,W}$ focus on location-specific visit requirements, allowing each location above the threshold to be a candidate home/work location. 
    Specifically, fractions of visits to home/work location candidates are computed relative to the total hours in which an individual has recorded data during the specified period.}
    \label{tab:parameters}

    \rowcolors{2}{gray!15}{white}  
    \resizebox{\textwidth}{!}{ 
    \begin{tabular}{lcl}  
        \toprule
        \textbf{Name}           & \textbf{Usage} & \textbf{Description} \\ 
        \midrule
        $\Delta T_H$     & Home estimation    & Size of the memory window for home estimation \\ 
        $\Delta T_W$     & Work estimation    & Size of the memory window for work estimation\\ 
        $C_{hours}$          & Home/Work temporal coverage filter    & Min. fraction of night/business hourly-bins with data in a day\\ 
        $C_{days,H}$         & Home temporal coverage filter & Min. fraction of days with data in a window\\ 
        $C_{days,W}$         & Work temporal coverage filter   & Min. fraction of days with data in a window\\ 
        $f_{hours,H}$         & Home location selection              & Min. fraction of night hourly-bins a location should be visited \\ 
        $f_{hours,W}$         & Work location selection               & Min. fraction of business hourly-bins a location should be visited\\ 
        $f_{days,W}$         & Work location selection            & Min. fraction of days a location should be visited within the window\\ 
        \bottomrule
    \end{tabular}
    }
\end{table}

For details on parameter implementation,  we refer to the Supplementary material Section S-8, where we provide the algorithm’s pseudocode.

\section{Results}\label{sec:results}
Here, we evaluate the performance of \textit{HoWDe} against ground truth data, explore the effect of parameter choices, and assess its robustness across demographic groups. 
Then, we demonstrate how  \textit{HoWDe} can be applied to estimate employment rates and commuting patterns.

\subsection{Algorithm performance: accuracy, retention, and robustness}\label{sec:valdation}

To assess how well \textit{HoWDe} can capture individuals' home and work locations, we measure i) the detected accuracy of the algorithm as the fraction of home/work locations that are correctly identified, and ii) the fraction of not-detected users, measured as the number of users where a home/work location was not detected against the number of users for which a home/work location was reported in the ground truth data (see Sec.~\ref{sec:methods-performance} for more details). 
We run the validation using the ground truth datasets D1 and D2, described in Methods.
In dataset D2, annotators reviewed complete stop sequences; therefore, the temporal dimension (changes of home and work locations) can not be considered.
For this reason, for D2, we consider a single window covering the entire data period (see Supplementary material Sec. S-9 for parameter details).

\subsubsection{Algorithm performance across parameter choices}\label{sec:expconfig}

We evaluate how the choice of configuration parameters affects \textit{HoWDe}’s performance. 
Among all parameters, the sliding-window length $\Delta_T$ and the minimum home-time fraction $f_{hours}$ have the strongest influence on home-detection accuracy, while other parameters exert more moderate effects. 
Temporal coverage filters, in particular, primarily determine the amount of usable data and should therefore be tuned according to the quality and completeness of the dataset at hand. 
A detailed analysis of these additional parameter configurations, including their dependence on the datasets' sparsity, is provided in Supplementary Material S-10.

\begin{figure}[!ht]
  \centering
  \includegraphics[width=1\textwidth]{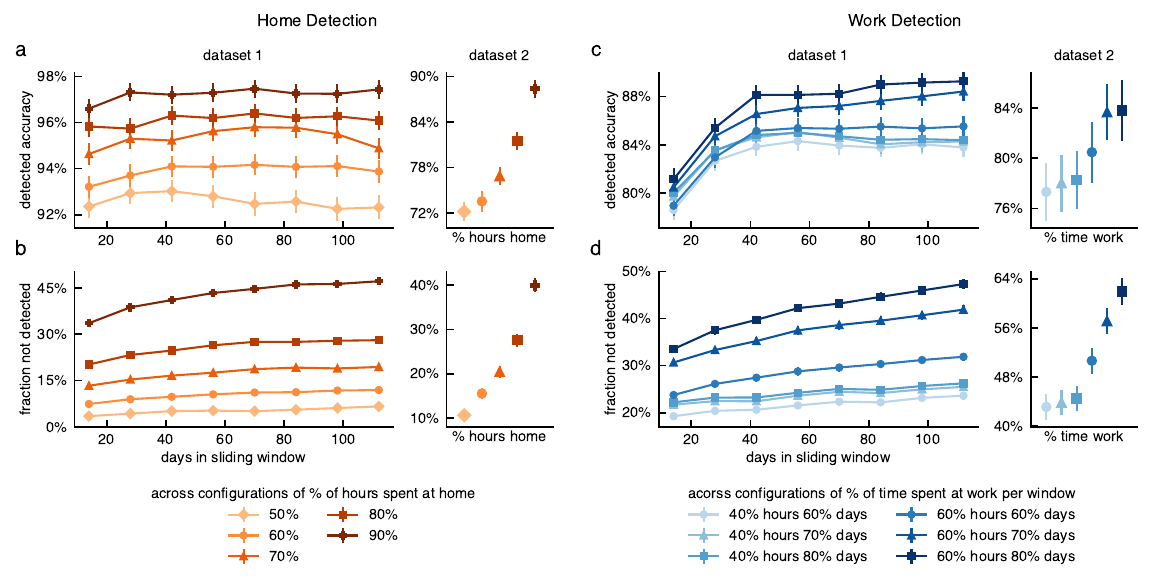}
  \caption{\small \textbf{Home and Work validation across configurations}
    a. Detected accuracy of the home location detection for datasets D1 (first column) and D2 (second column). Each marker represents different configurations of the minimum fraction of hours an individual must be at home per night ($f_{hours, H}$). 
    For dataset D1, we explore how the results change for each sliding window increase (x-axis), while for dataset D2, we focus on a single sliding window (comprising the entire data period). 
    b. Fraction of not detected home locations for dataset D1 (first column) and dataset D2 (second column), sharing with the panel a) the legend.
    c. Detected accuracy of the work location detection for datasets D1 (first column) and D2 (second column). Each marker represents different configurations of the minimum fraction of hours an individual must be at work within the typical business hours ($f_{hours, W}$), and the minimum fraction of days an individual should visit the potential work across weekdays ($f_{days, W}$). For dataset D1, we explore how results change with increasing sliding window sizes, while for dataset D2, we keep the same sliding window size (covering the entire period). 
    d. Fraction of not detected work locations for dataset D1 (first column) and dataset D2 (second column). The legend is shared with panel c.
    }
    \label{fig:home_work_config}
\end{figure}

Both the detected accuracy of the home location estimation and the fraction of non-detected homes increase as the minimum home-time fraction threshold $f_{hours, H}$ increases (Figure~\ref{fig:home_work_config}a,b). 
In dataset D1, the fraction of non-detected homes increases with the window size, but the detected accuracy shows no substantial improvement beyond $\Delta_T = 28$ days ($t \pm 2$ weeks).
This suggests that a window of $4$ weeks offers a good balance between high accuracy and user retention for home detection.

Table~\ref{tab:accuracy} reports the results by focusing on two configuration setups: the minimum configuration, which maximises detected accuracy while minimising the number of non-detected users, and the maximum configuration, which prioritises maximum detected accuracy at the cost of higher user non-detection rates.
We also report the results obtained by two state-of-the-art home/work detection algorithms: \textit{Atlas} \cite{TheAtlasofInequality,moro2021mobility} and \textit{TimeGeo} \cite{jiang2016timegeo} (see Secs.~\ref{sec:lit_rev} and \ref{sec:methods-comparison} for algorithm and performance comparison details).

The maximum configuration is reached for $\Delta_{T_H}= 28$ days window and $90\%$ minimum home-time fraction (see Supplementary material Sec. S-9 for parameter details). 
With this choice of parameters, the algorithm achieves a detected accuracy of $97.30\% \pm 0.39\%$ for D1, and $88.47 \%\pm 1.13\%$ for D2, but comes at the cost of excluding nearly two out of every five users (see Table~\ref{tab:accuracy}).
The minimum configuration is obtained for parameters $\Delta_{T_H} = 28$ days window and $50\%$ home frequency (see Supplementary material Sec.~S-9 for parameter details). 
Here, the algorithm achieves $92.94\%\pm 0.49\%$ detected accuracy for D1, and $72.34\% \pm 1.13\%$ for D2, while maintaining a fraction of non-detected of $4.29\% \pm 0.38\%$ for D1 and  $10.44\% \pm 1.02\%$ for D2.

All in all, we find that \textit{HoWDe} achieves higher detected accuracy than previous heuristic methods \cite{jiang2016timegeo, moro2021mobility}, partly because it excludes users with low-quality data (see Table~\ref{tab:accuracy}).

Shifting to work location detection, we find that the fraction of correctly detected work locations increases with the window size until $\Delta_{T_W}= 42$ days ($t \pm 3$ weeks), 
after which detected accuracy stabilizes while non-detection rates continue to rise (see Figure~\ref{fig:home_work_config}c,d). 

In the work location detection, two key parameters are the minimum fraction of hourly bins $f_{hours,W}$ and days ${f_{days,W}}$ required for a location to be selected as work. 
Examining the impact of these two parameters (see Figure~\ref{fig:home_work_config}c-d, different colors), we find that stricter thresholds enhance detected accuracy in D1 but increase non-detection rates. 
For a larger value of ${f_{hours,W}}=60\%$, increasing ${f_{days,W}}$ enables significant accuracy gains. 
In contrast, for ${f_{hours,W}}=40\%$, changes in ${f_{days,W}}$ have a weaker effect. 
We find similar results for D2, when changing the minimum fraction of days ${f_{hours,W}}$ and fraction of hours ${f_{hours,W}}$.

\textit{HoWDe} achieves an average work detection accuracy of $83.84\% \pm 0.80\%$ and $88.14\% \pm 0.83\%$ for the minimum and maximum configurations in dataset D1 (see Supplementary material Sec. S-7 for parameter details), outperforming baseline methods (see Table~\ref{tab:accuracy}).
Similarly, for D2, \textit{HoWDe} achieves high detected accuracy levels ($81.86\%\pm2.13\%-74.43\%\pm2.13\%$), but, also in this case, systematically lower than D1 (see Table~\ref{tab:accuracy}). 
This difference in performance may be partly explained by the fact that dataset D2 was collected during the COVID-19 pandemic, when people did not follow traditional mobility patterns. 

While \textit{Atlas} does not explicitly estimate work locations, our adaptation (see Sec.~\ref{sec:lit_rev}) outperforms \textit{TimeGeo}.
This highlights that methods based solely on distance and visit counts, such as \textit{TimeGeo}, perform substantially worse than those incorporating visit duration, including our \textit{Atlas} adaptation and \textit{HoWDe}. By accounting for missing data within individuals, \textit{HoWDe} achieves the highest overall detected accuracy.
\input{table_1}

\subsubsection{Algorithm robustness across demographic groups}\label{subsec:res-demo}

\begin{figure}[!ht]
        \centering
        \includegraphics[width=\linewidth]{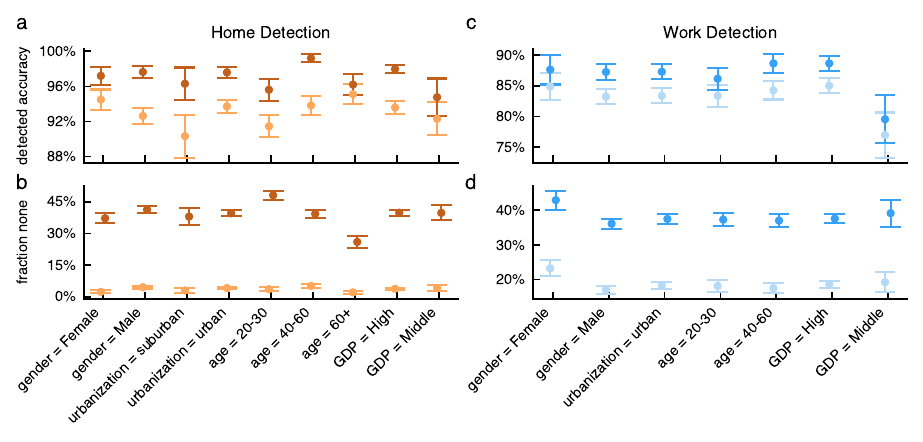}
        \caption{\small 
        \textbf{Home and work detection accuracy and fraction of not detected locations across demographics for dataset D1.} 
        a. Detected accuracy of home location detection across demographic groups. Dark orange illustrates the results for the maximum configuration, while light orange illustrates the minimum. Error bars indicate bootstrapped error estimates (applies to all panels).
        b. Fraction of not detected home locations (configurations color coding is shared with panel a). 
        c. Detected accuracy of work location detection across demographic groups. Dark blue illustrates the results for the maximum configuration, while light blue illustrates the minimum.
        d. Fraction of not detected work locations  (configurations color coding is shared with panel c). 
        }
        \label{fig:acc_by_demo}
\end{figure}

\textit{HoWDe}'s accuracy comes at the cost of user retention. 
The exclusion of users may introduce biases toward specific demographic groups.
To assess these biases, we measure detected accuracy and user retention for different groups based on age (self-reported), gender (self-reported), urbanisation levels (estimated using the 2022 GHS settlement model grid at 1 km resolution~\cite{schiavina2022ghsl}), and country GDP.
Results are obtained for dataset D1, using the minimum and maximum parameter configurations (see Figure~\ref{fig:acc_by_demo}), and explored only for subgroups comprising at least 200 users.

The largest detected accuracy gaps appear in suburban areas and middle-income countries. 
In suburban settings, home detection accuracy drops by $3.39$ points compared to urban areas under the minimum configuration. However, user retention remains stable, suggesting that the decline in detected accuracy is not due to a disproportionate loss of suburban users.

In middle-income countries, home detection accuracy falls only under the maximum configuration ($-3.31$ points), while work detection exhibits larger drops of $9.13$ and $8.06$ points under the maximum and minimum configurations, respectively. 
These declines occur without changes in user retention, suggesting that the reduced detected accuracy reflects differences in work visitation patterns rather than sample loss.

Across other demographic groups, we find significant differences in home detection. 
Home detected accuracy drops by $1.87$ points for males compared to females under the minimum configuration. 
For individuals aged 20–30, detected accuracy decreases by $2.40$ and $3.60$ points under the minimum and maximum configurations, respectively, relative to the 40–60 age group. 
 
Despite accuracy deviations, home detection accuracy remains within $90\%-99\%$ and work detection within $77\%-89\%$ across all lower-accuracy subgroups (see Figure~\ref{fig:acc_by_demo}), aligning with the observed range for D1 and D2.

These findings emphasize that, while detection accuracy remains consistent across demographic groups and aligns with overall results, it is essential to validate sample representativeness when applying home and work detection algorithms.

\subsection{Impact of parameter sensitivity on real-world applications}\label{sec:implications}
Small detection errors can propagate across applications, potentially undermining the reliability of downstream results.
To address this potential issue, we examine how tuning the detection algorithm parameters impacts results in two downstream applications: (i) employment rate estimates in Europe and (ii) commuting pattern characteristics.

\subsubsection{Estimating regional employment rates}
We estimate the regional employment rate by measuring the fraction of individuals with a detected work location, stable for a minimum of $30$ days. 
We then investigate how the employment estimates obtained from a ``minimum'' and ``maximum'' configurations differ and how much they deviate from official regional statistics~\cite{LabourMarketStats2024Eurostat}. 
This analysis covers $90$ regions across six European countries, each with a minimum of $1000$ individuals per region.

Figure \ref{fig:case_study}.a shows that both configurations produce similar geographical patterns in regional employment. 
However, while both exhibit moderate to high Pearson correlation, the maximum configuration has a stronger correlation with reported employment rates but estimates fewer regions due to stricter data coverage thresholds. 
It also has a higher overall relative error (Figure \ref{fig:case_study}.b, and Supplementary material Figure S-18, except in Germany, where the trend reverses. This suggests that while the maximum configuration enhances individual accuracy, the minimum configuration may be more suitable for population estimates.

\subsubsection{Home-Work Commuting distances}
This case study examines whether commuting distance, measured as the great-circle distance between detected home and work locations, varies with population density and whether parameter configurations can influence results. 
We estimate urbanisation levels and population density around users’ home locations using the 2022 GHS settlement model grid at 1 km resolution~\cite{schiavina2022ghsl}.

\begin{figure}[!ht]
    \centering
    \includegraphics[width=\linewidth]{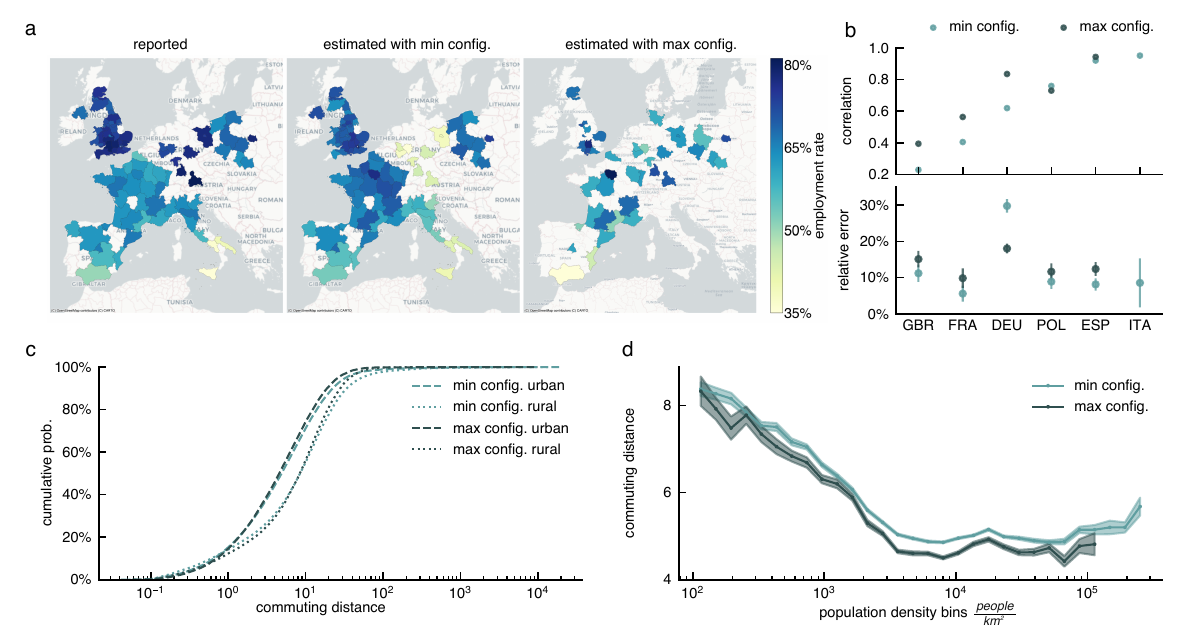}
    \caption{\small \textbf{Applications for home and work detection}
    a. Employment rate by province, as reported (left), measured using the configuration with the minimum user loss (middle), and measured with the configuration with the highest detected accuracy (right).
    b. Difference between the minimum (light teal) and maximum configurations (dark teal) by country, as the correlation between the reported and selected configuration (top), the relative error (bottom).
    c. Difference in commuting distance (km) between urban (dashed) and rural (dotted) by configuration. 
    d. Difference in commuting distance (km) against the median and standard error of the population density bins ($\frac{people}{km^2}$).
    }
    \label{fig:case_study}
\end{figure}

Our analysis reveals consistent findings across configurations: commuting distances are longer in rural areas compared to urban areas, with an absolute difference of $3.61\pm 0.04$ km for the ``minimum'' and $3.85\pm 0.06$ km for the ``maximum'' configurations (see Figure \ref{fig:case_study}.c).
Additionally, commuting distances decrease with increasing population density (see Figure \ref{fig:case_study}.d).
However, the maximum configuration results in shorter commuting distances, with the most densely populated areas showing the largest differences between configurations.
This highlights the need for robustness checks to ensure consistency across different configurations.

\section{Discussion}
Home and work location detection from smartphone data is a widely used task in mobility and transport research. 
Despite its central role in many studies, most existing approaches have not been systematically validated against ground truth and do not explicitly account for the heterogeneous and irregular patterns of missing data that characterise real-world smartphone datasets.

In this work, we proposed \textit{HoWDe}, an algorithm for home and work location detection from stop-location data.
\textit{HoWDe} integrates key features of state-of-the-art heuristic approaches and improves robustness against missing data by allowing for heterogeneous data gaps within and across users. 
\textit{HoWDe} builds on insights derived from empirical home-work visit data. 
Using two datasets capturing stop-locations and ground truth home and work locations (D1, D2) for $5,099$ individuals from $68$ countries, we characterised home/work visit patterns. 
We found that individuals are most likely at home between midnight and 6 a.m. ($\sim75\%$ probability) and at work between 9 a.m. and 4 p.m. on workdays ($\sim50\%$). 
However, visit patterns and missing data vary widely across users and over time, and even within individual users’ routines.

We validated \textit{HoWDe} against the ground truth datasets, achieving up to $\sim97\%$ detected accuracy in D1 and $\sim88\%$ in D2 for home detection, and up to $\sim88\%$ detected accuracy in D1 and $\sim82\%$ detected accuracy in D2 for work detection. 
We found that \textit{Atlas} and \textit{Timegeo} achieve lower overall detected accuracy than \textit{HoWDe}. 
However, they attain high home-detection performance ($81.44\% - 82.92\%$ for D1 and $70.10\% - 70.62\%$ for D2), indicating that both are adequate for this task. 
In contrast, for work detection, relying solely on distance and visit counts—as in \textit{Timegeo}—yields much lower detection accuracy ($37.51\%$ for D1 and $56.89\%$ for D2), whereas incorporating duration, as in our adaptation of \textit{Atlas} and in \textit{HoWDe}, results in markedly higher accuracy.

Our analysis highlights the trade-off between achieving accuracy and retaining users with low-quality data. 
We identified two key parameters affecting performance: the window length considered ($\Delta_T$) and the minimum visit proportion required for classification ($f_{hours, H,W}$, $f_{days,W}$).
A window of at least $28$ days ensures a reliable home estimation, while work windows longer than $42$ days reduce user retention without significant accuracy gains.

We also revealed how the algorithm's detected accuracy varies across population groups, with the largest gaps appearing between urban and suburban areas (for home detection), and between high- and middle-income countries (for work detection).
Although these accuracy gaps persist across configurations, two case studies show that prioritizing accuracy over user retention is preferable for individual-level tasks, whereas maximizing retention may be more valuable for population-level analyses.

The insights gained from this study, along with a formalized algorithm, can support research on commuting~\cite{lucchini2023socioeconomic,Centellegher2025}, social contact networks~\cite{aleta2022quantifying}, and access to amenities~\cite{lucchini2021living}, with implications for areas such as computational social science, urban planning, and public health~\cite{oliver2020mobile,barreras2024exciting}.
Improving location data processing will enhance our understanding of both universal mobility laws and micro-level social dynamics~\cite{schlapfer2021universal}.  

Despite its strengths, \textit{HoWDe} has key limitations. 
While providing a robust framework for detecting anchor locations, it does not infer their semantic purpose---that is, it cannot distinguish between work, study, or other activities with similar temporal patterns.
Future work could address this by integrating contextual data, such as point-of-interest (POI) information or survey-based semantic labels. 
In addition, \textit{HoWDe} assumes a single dominant lifestyle pattern across users and therefore cannot, within a single run, distinguish between regular schedules and rotating night-shift routines.  While this assumption aligns with the empirical regularities in our ground truth data, incorporating an initial lifestyle-detection step represents a promising direction for future work.  
Finally, the current implementation identifies a single home and work location per day and does not explore multi-home or multi-work scenarios. Leveraging the ranked lists of potential home and work locations (stored internally in \textit{HoWDe}) could support such analyses, but this lies beyond the evaluation possibilities afforded by our ground truth data.
More advanced methods may also be needed to handle heterogeneous GPS data and ensure its applicability across contexts, in particular for rural and less densely populated areas~\cite{rinzivillo2014purpose,gvirtz2024limits}.

Working with individual-level location data presents well-known challenges for data sharing and reproducibility due to the high risk of re-identification~\cite{de2013unique,zang2011anonymization}. Under the EU General Data Protection Regulation, re-identification risk arises when: (i) the data is not k-anonymous—that is, individual trajectories are unique—and (ii) the data can be linked, through means that are reasonably likely to be used, to an external dataset containing direct personal identifiers (GDPR Recital 26). 
The first condition is almost always met in human mobility datasets: even after coarse-graining or obfuscating data points, trajectories remain highly idiosyncratic~\cite{de2013unique}, making k-anonymity infeasible without rendering the data analytically useless. 
The second condition can be mitigated through privacy-preserving techniques—such as perturbing spatial coordinates, obfuscating timestamps, or omitting sensitive fields—to increase the difficulty of linking mobility traces to known individuals. 
These transformations can in some cases reduce re-identification risk to a negligible level, but often at the cost of limiting the usefulness of the data for certain analytical purposes. 
This creates a fundamental trade-off between privacy protection and scientific utility, which is why access to individual-level location data is typically restricted to secure environments and regulated by data-use agreements.

Because of these constraints, many data holders perform sensitive inference steps—such as detecting home and work locations—internally. 
However, the methods used for this purpose are often proprietary or undocumented, introducing uncertainties and reducing reproducibility. 
Our contribution directly addresses this gap by providing a validated, open-source pipeline for detecting home and work locations that can be executed in-house by data holders prior to data sharing. 
This supports methodological transparency and consistency, while enabling the release of privacy-preserving data products—such as aggregated visitation patterns or synthetic traces.

In summary, by advancing methods for home and work detection, this study contributes to a more robust and scalable foundation for mobility research. 
By providing an open-source, validated solution, we hope to facilitate more reliable and ethical research, contributing to a deeper understanding of human mobility and its implications for society.

\section{Code and Data Availability}\label{sec:data_code_avail}
All code needed to implement \textit{HoWDe} is available as a \href{https://github.com/LLucchini/HoWDe/tree/main}{GitHub repository} at \cite{gitRepoCode}, along with usage examples in the GitHub folder \textit{/tutorials}. 
We additionally provide \textit{HoWDe} as a Python package on PyPI for direct installation via \texttt{pip}.

Data to ensure replicability of the analysis performed in this work can be found at \\\href{https://doi.org/10.11583/DTU.28846325}{https://doi.org/10.11583/DTU.28846325}.
To protect user anonymity, we applied the following privacy-preserving transformations to the stop sequences. 
First, time was discretised into 10-minute intervals.
Second, each user’s timeline was shifted so that their first recorded activity occurred in the first week of 1970.
Third, entire days were randomly shuffled within each week, while preserving the distinction between weekdays and weekends.
All data analysis was carried out in accordance with the European Union's General Data Protection Regulation 2016/679 (GDPR).

\newpage
\bibliographystyle{unsrt} 


\end{document}

%% file: table_1.tex
\begin{table}[H]
\centering
\renewcommand{\arraystretch}{1.2} 
\setlength{\tabcolsep}{8pt}       
\resizebox{0.9\textwidth}{!}{%
    \begin{tabular}{llllll}
        \toprule
        \textbf{Detected accuracy} &   
        & \textbf{HoWDe min} & \textbf{HoWDe max} & \textbf{Atlas} & \textbf{TimeGeo} \\
        \cline{1-6}
        \multirow[t]{2}{*}{\textbf{Home}} & D1 &
        $\mathbf{92.94 \pm 0.49}$ &
        $\mathbf{97.30 \pm 0.39}$ &
        $82.92 \pm 0.66$ &
        $81.44 \pm 0.70$\\
        & D2 &
        $\mathbf{72.34 \pm 1.13}$&
        $\mathbf{88.47 \pm 1.13}$&
        $70.62 \pm 1.06$& 
        $70.54 \pm 1.15$\\
        \cline{1-6}
        \multirow[t]{2}{*}{\textbf{Work}} & D1 &
        $\mathbf{83.84 \pm 0.80}$ &
        $\mathbf{88.14 \pm 0.83}$ &
        $62.20 \pm 0.92$ &
        $37.51 \pm 0.92$ \\
         & D2 &
         $\mathbf{76.43 \pm 2.13}$&
         $\mathbf{81.86 \pm 2.18}$&
         $71.74 \pm 2.17$& 
         $56.93 \pm 2.35$\\
        \toprule
        & &  &  &  &  \\
        \toprule
         \textbf{Fraction not detected} &   
         & \textbf{HoWDe min} & \textbf{HoWDe max} & \textbf{Atlas} & \textbf{TimeGeo} \\
        \cline{1-6}
        \multirow[t]{2}{*}{\textbf{Home}} & D1 &
        $4.29 \pm 0.38$ &
        $38.75 \pm 0.93$ &
        $\mathbf{0.0\pm 0.0}$ & 
        $\mathbf{0.0 \pm 0.0}$ \\
        & D2 &
        $10.44 \pm 1.02$&
        $39.94 \pm 1.43$&
        $\mathbf{0.0 \pm 0.0}$ & 
        $\mathbf{0.0 \pm 0.0}$ \\
        \cline{1-6}
        \multirow[t]{2}{*}{\textbf{Work}} & D1 &
        $20.64 \pm 0.77$ &
        $39.71 \pm 0.93$ &
        $\mathbf{0.0 \pm 0.0}$&
        $0.80 \pm 0.16$ \\
         & D2 &
         $34.40 \pm 1.93$&
         $58.73 \pm 1.87$&
         $\mathbf{0.0 \pm 0.0}$ &
         $0.68 \pm 0.36$\\
        \bottomrule
    \end{tabular}
    }
\caption{\textbf{Performance comparison of home and work detection methods.} Detected accuracy (top) and fraction not detected (bottom) across datasets D1 and D2. Bold highlights indicate the best method per metric—highest accuracy or lowest missing rate. While Atlas does not explicitly detect work locations, we adapt its home location algorithm for this purpose (see Section~\ref{sec:lit_rev}).}
\label{tab:accuracy}
\end{table}